\documentstyle[12pt]{article}
\newcommand{\vect}[1]{\mbox{\bf #1}}
\newcommand{\vecs}[1]{\mbox{\scriptsize \bf #1}}
\newcommand{\rms}[1]{\mbox{\scriptsize #1}}
\newcommand{\rmt}[1]{\mbox{\tiny #1}}

\newcommand{\lw}[1]{\smash{\lower 2ex\hbox{#1}}}

\begin{document}
%
%
\title{{\Large\bf Coulomb wave function correction
including}\vspace{-4mm}\\
{\Large\bf momentum resolution for charged hadron
pairs:}\vspace{-4mm}\\
{\large\bf Analysis of data for $\pi^+\pi^-$ pair production in $p\;+$ Ta
reaction}\vspace{-4mm}\\
{\large\bf at 70 GeV/c}}
\author{T.~Mizoguchi$^1$\thanks{e-mail: mizoguti@yukawa.kyoto-u.ac.jp},~
M.~Biyajima$^2$\thanks{e-mail: minoru44@jpnyitp.bitnet},~
I.~V.~Andreev$^3$\thanks{e-mail: andreev@lpi.ac.ru}~~and~
G.~Wilk$^4$\thanks{e-mail: wilk@fuw.edu.pl}\\
{\small\sl $^1$Toba National College of Maritime Technology, Toba
517, Japan}\vspace{-2mm}\\
{\small\sl $^2$Department of Physics, Faculty of
Science,}\vspace{-4mm}\\
{\small\sl Shinshu University, Matsumoto 390, Japan}\vspace{-2mm}\\
{\small\sl $^3$P.~N.~Lebedev Institute of Physics (FIAN), 117 924
Moscow, Russia}\vspace{-2mm}\\
{\small\sl $^4$Soltan Institute for Nuclear Studies, Nuclear Theory
Department}\vspace{-4mm}\\
{\small\sl Ho\.za 69, PL-00-681 Warsaw, Poland}
}
\date{{\small Preprint DPSU-96-9 \qquad and \qquad SINS-PVIII-1996-15}\\
\today}
\maketitle
%
%
\begin{abstract}
We propose a new method for the Coulomb wave function correction
including the momentum resolution for charged hadron pairs and
apply it to the precise data on $\pi^+\pi^-$ correlations obtained in
$p + \mbox{Ta}$ reaction at 70 GeV/c. It is found that interaction
regions of this reaction (assuming Gaussian source function) are $5.6
\pm 3.0$ and $4.4 \pm 2.6$ fm for the thicknesses of the target $8$
and $1.4$ microns, respectively. The physical picture of the source
size obtained in this way is discussed. 
\end{abstract}
\thispagestyle{empty}
\newpage
\setcounter{page}{1}
%
%
\noindent
{\bf 1. Introduction.}\quad Recently we have obtained the new
formulae for the Coulomb wave function correction for charged hadron 
pairs~\cite{biya95,biya96}. In particular we have applied them (in
\cite{biya96}) to data on $\pi^+\pi^-$ correlation obtained in $p +
\mbox{Ta}$ reaction at 70 GeV/c~\cite{afan90} (which were originally
corrected by usual Gamow factor only). However, as it was pointed out
to us by one of the author of \cite{afan90}, we did not take into
account their published finite momentum resolution \cite{afan95}. In
fact, our formulae cannot be applied directly to experimental data
in which such momentum resolution is accounted for. Therefore in
the present letter we would like to extend our method for the Coulomb
wave correction provided in \cite{biya95,biya96} to include also the
momentum resolution case and to re-analyse data of \cite{afan90} and
also to analyse the new, preliminary data of \cite{afan96} obtained
with two kinds of thickness of Ta target: 8 microns (8 mkm) and 1.4
microns (1.4 mkm).\\ 

In the next paragraph we first reconstruct (for the sake of
completeness) the analysis performed in Ref.~\cite{afan90} and then, 
in the third paragraph,  we propose our new method for the Coulomb
wave correction including this time also the momentum resolution. The
final part contains our concluding remarks.\\ 

%
%
\noindent
{\bf 2. Reconstruction of the analysis of $\pi^+\pi^-$ correlation
data performed by using the Gamow factor with momentum
resolution.}\quad It has been stressed in~\cite{afan90} that
relative momenta of $\pi^+\pi^-$ pairs observed by them have some
finite resolutions. The averaged correlation function, defined as
\begin{equation}
  R(\pi^+\pi^-) = \left\langle \frac 1{\sigma}
\frac{d^2\sigma}{dp_1dp_2} \left/ \left(\frac 1{\sigma}
\frac{d\sigma}{dp_1}\frac 1{\sigma} \frac{d\sigma}{dp_2}(1 + B_{pr})
\right)\right.\right\rangle ,
  \label{eqn:a1}
\end{equation}
depends therefore on this momentum resolution, where $(1 + B_{pr})$ 
stands for non-Coulomb correlation factor. To account for it the
following random number method has been proposed in Ref.
\cite{afan90} in order to obtain the corresponding averaged
quantities in analysis of the correlation data.\\

First of all, the relative momentum of the measured pair, $q = p_1 -
p_2$, is decomposed into its longitudinal and transverse components:
$q_{\rmt L}$ and $q_{\rmt T}$, respectively, by making use of the
uniform random number $u\in (0,1)$ (it is worthwhile to notice at
this point that the transverse components $q_{\rmt T}$ in data of
\cite{afan90} are smaller than $10$ MeV/c). One uses the following
scheme here: 
\begin{eqnarray}
  & & q_{\rmt T} = \left\{ 
  \begin{array}{l}
    10\sqrt u \quad\mbox{for}\quad q \ge 10\mbox{MeV/c},\; , \\
     q\sqrt u \quad\mbox{for}\quad q \le 10\mbox{MeV/c}\; , 
  \end{array}\right. \label{eq:a2}\\
  & & q_{\rmt L} = \sqrt{q^2 - q_{\rmt T}^2}\:.\nonumber
\end{eqnarray}

At the next step, the Gaussian random numbers for $q_{\rmt L}$ and
$q_{\rmt T}$ are generated in the following way:
\begin{eqnarray}
  q_{\rmt L\rms{(random)}} &=& \sigma_{\rmt L} X + q_{\rmt L}\:,
  \label{eq:a3}\\
  q_{\rmt T\rms{(random)}} &=& \sigma_{\rmt T} X + q_{\rmt T} .
  \label{eq:a4}
\end{eqnarray}
In the above equations $X$ stands for the standard Gaussian random
number\footnote{
The probability density of such random numbers is given by
$$
f(x) = \frac 1{\sqrt{2\pi}} \exp(-x^2/2) \quad (-\infty < x <
\infty)\:.
$$
}
whereas $\sigma_{\rmt L}$ and $\sigma_{\rmt T}$ are longitudinal and
transverse setup resolutions for the corresponding components, which
are equal to (values used in~\cite{afan96}): $\sigma_{\rmt L} = 1.3$
MeV/c; $\sigma_{\rmt T} = 0.6$ MeV/c (for target of the thickness $8$
mkm) and $\sigma_{\rmt T} = 0.4$ MeV/c (for the 1.4 mkm target).\\

Finally, using the randomized number $q_{\rms{random}} =
\sqrt{q_{\rmt L\rms{(random)}}^2 + q_{\rmt T\rms{(random)}}^2}$ one
calculates the corresponding randomized Gamow factor correction:
\begin{equation}
  G(-\eta_{\rms{random}}) =
\frac{-2\pi\eta_{\rms{random}}}{\exp(-2\pi\eta_{\rms{random}}) -
1}\:,
  \label{eqn:a5}
\end{equation}
where $\eta_{\rms{random}} = m\alpha/q_{\rms{random}}$. The full flow
chart for this procedure is shown in Fig.~1. Calculating now the
average value of $G(-\eta_{\rms{random}})$ in 100 k events one
can estimate the Gamow factor with this finite momentum resolution,
\begin{equation}
  R(q) = \widetilde{G}(-\eta)b + (1-b)\:,
  \label{eqn:a6}
\end{equation}
where $b$ is a free parameter. It is understood (or, rather,
implicitly assumed) that essentially all unlike sign pions one deals
with here originate from decays of long lived particles like $\eta$,
$K_0^S$, $\Lambda$, and so on\footnote{This fact has some
consequences on the determination of the radius of the interaction
region as will be shown later on and makes it different from that
obtained from the Bose-Einstein correlations.}. Figure~2 shows the
results of analysis of new data (for 8 mkm target)~\cite{afan96} for
region $q > 3$ MeV/c using this method.\\  

%
%
\noindent
{\bf 3. Proposal of the new method.}\quad We would like to propose
now a new method of Coulomb wave function correction with a source
function $\rho(r)$ instead of the Gamow factor, in order to analyse
the  same data. As usual we decompose the wave function of unlike
charged bosons with momenta $p_1$ and $p_2$ into the wave function of
the center-of-mass system (c.m.) with total momentum $P =
\frac{1}{2}(p_1 + p_2)$ and the inner wave function with relative
momentum $q = (p_1 - p_2)$. This allows us to express Coulomb wave
function $\Psi(\vect{q},\vect{r})$ in terms of the confluent
hypergeometric function $\Phi$~\cite{schiff}: 
\begin{equation}
 \Psi(\vect{q},\vect{r}) = \Gamma(1-i\eta)e^{\pi \eta/2}
e^{i\vecs{q}\cdot\vecs{r}/2} \Phi(i\eta;1;iqr(1 - \cos
\theta)/2)\:,\\
  \label{eqn:a7}
\end{equation}
where $r = x_1 - x_2$ and the parameter $\eta = m\alpha/q$. Assuming
factorization in the source functions, $\rho(r_1,r_2) =
\rho(r_1)\rho(r_2) = \rho(R)\rho(r)$ (here $R = \frac{1}{2}(x_1 +
x_2)$), we obtain the expression for Coulomb correction for the
system of $\pi^+\pi^-$ pairs identical (modulo the sign) as
in~\cite{bowler91,biya95,biya96}:
\begin{eqnarray}
  C_{\rmt C}(-\eta) &=& \int \rho(R) d^3R \int \rho(r) d^3r
|\Psi(\vect{q},\vect{r})|^2 \nonumber\\
  &=& G(-\eta)\sum_{n=0}^{\infty} \sum_{m=0}^{\infty}
\frac{(-i)^n(i)^m}{n+m+1}\, q^{n+m}\, I_{\rmt R}(n,m) A_n A_m^*
\nonumber\\
  &=& G(-\eta)[1 + \Delta_{1\rmt C}(-\eta)]\:,
  \label{eqn:a8}
\end{eqnarray}
where
$$
I_{\rmt R}(n,m) = 4 \pi \int dr\, r^{2 + n + m} \rho (r),\qquad A_n =
\frac{\Gamma(-i\eta + n)}{\Gamma(-i\eta)}\frac{1}{(n!)^2}\:.
$$
For the specific choice of Gaussian source distribution, $\rho(r) =
\frac{\beta^3}{\sqrt{\pi^3}}\exp(-\beta^2 r^2)$, we have
\begin{equation}
  I_{\rmt R}^{\rmt G} (n,m) = \frac 2{\sqrt{\pi}} \left(\frac
1{\beta}\right)^{n+m} \Gamma\left(\frac{n+m+3}2\right) ,
  \label{eqn:b1}
\end{equation}
whereas exponential source function, $\rho(r) = \frac{\beta^3}{8\pi}
\exp(-\beta r)$, leads to
\begin{equation}
  I_{\rmt R}^{\rmt E}(n,m) = \left(\frac 1{\beta}\right)^{n+m}
\frac{(n+m+2)\,!}2\:.
  \label{eqn:b2}
\end{equation}
Using now the same method of Gaussian random numbers as in the
previous paragraph, we can analyse the old and the new data on
$\pi^+\pi^-$ pairs~\cite{afan90,afan96} using the following formula: 
\begin{equation}
  R(q) = \widetilde{C}_{\rmt C}(-\eta)b + (1-b)\:.
  \label{eqn:a9}
\end{equation}
Figs.~3 and 4 show results of our analysis of the old and the new
data, respectively. Table~\ref{tbl:a1} show our results obtained
using eq.~(\ref{eqn:a8}) applied to old and new data with $q > 3
$MeV/c.\\

%
%
\noindent
{\bf 4. Concluding remarks.}\quad We have proposed the new method for
the Coulomb wave function correction with momentum resolution and
applied it to the analysis of the precise data provided
by~\cite{afan90,afan96}. Authors of Ref.~\cite{afan90} have 
analysed their $\pi^+\pi^-$ correlation data using Gamow factor for
Coulomb corrections together with the random numbers method to
account for final momentum resolution. We have repeated this analysis
replacing Gamow factor by the Coulomb wave function but following the
same method for correction for the momentum resolution effect (cf.
eq.~(\ref{eqn:a8})). As a result we have obtained the following
ranges of interaction for the Gaussian source function:
\begin{eqnarray}
  r(p + \mbox{Ta}) = \frac 1{2\beta} &=& 5.6\pm 3.0 \mbox{ fm\quad
for 8 mkm},\\
  &=& 4.4\pm 2.6 \mbox{ fm\quad for 1.4 mkm}.\nonumber
\end{eqnarray}
To get a correct physical picture of the source size, we should
calculate the root mean squared size, which is equal to:
\begin{eqnarray}
  r_{\rms{rms}} = \frac{\sqrt 3}{2\beta} &=& 9.7\pm 5.3 \mbox{
fm\quad for 8 mkm},
  \label{eqn:a10}\\
  &=& 7.7\pm 4.5 \mbox{ fm\quad for 1.4 mkm}.\nonumber
\end{eqnarray}
The present study of $\pi^+\pi^-$ pair correlations has shown
therefore that one can estimate the interaction region even from the
$\pi^+\pi^-$ correlation data. It can be compared with the size of
the Ta nucleus, which is given by: 
\begin{eqnarray}
  r_{\rmt{Ta}} &=& 1.2\times A^{1/3}\nonumber\\
              &=& 1.2\times (181)^{1/3} = 6.8 \mbox{ fm} .
  \label{eqn:a12}
\end{eqnarray}
As one can see, $r_{\rms{rms}}$ is significantly bigger than
$r_{\rmt{Ta}}$. We attribute this difference to a physical picture 
shown in Fig.~5, i.e., to the fact that unlike-sign pions are mostly
(if not totally) emerging from the long-lived resonances shown there.
(In a future one should consider also a possibility of more direct
estimation of the parameter $b$ and its role in determining the
source size parameter\footnote{In an analysis of
$\pi^0\pi^0$ correlation data, a similar function $f(q)$ is introduced:
$$
R(\pi^0\pi^0) = f(q) + (1 - f(q))[1 + \lambda E_{2\rmt B}^2]\:,
$$
where $\lambda$ and $E_{2 \rmt B}$ are the degree of coherence and an
exchange function due to the Bose-Einstein effect. $f(q)$ is
attributed the resonances effect; $f(q) \approx 0.9\sim 0.7$ depends
on the Monte Carlo programs~\cite{wolf94}.}).\\

For completeness we have also tried to analyse the same data using
exponential source function instead of Gaussian. As is shown in
Table~\ref{tbl:a2} this leads to errors on $r=\frac{1}{2\beta}$ of
the order of $100$\%, i.e., with this type of source function we
cannot estimate the source size (therefore it has to be discarded).\\

%
%
{\bf Acknowledgements:}\quad Authors are grateful to Dr.
L.~G.~Afanas'ev for his kind correspondences and for providing us
with the new data on $\pi^+\pi^-$ correlation prior to publication.
This work is partially supported by the Grant-in-Aid for Scientific
Research from the Ministry of Education, Science and Culture of Japan
(\# 06640383), and the Japan Society of promotion of Science (JSPS).
One of authors (I.~A.) is also partially supported by Russian Fund of 
Fundamental Research (grant 96-02-16347a).
%
%

%
%
\newpage
\begin{table}[htbp]
\begin{center}
\caption[table1]{Results of the $\chi^2$ fits of $R(q)$ for Gaussian
source by eqs.~(\ref{eqn:a6}) and (\ref{eqn:a9}).}
\medskip
\label{tbl:a1}
\begin{tabular}{c|cccc}
\hline
Reaction & Formula & $1/2\beta$ [fm] & $b$ & $\chi^2/NDF$\\
\hline
data of Ref.~\cite{afan90} & Gamow factor & --- & --- &
57.8/40\\
cf. Ref.~\cite{biya96} 
& eq.~(\ref{eqn:a8}) & $2.30\pm 0.88$ 
& --- & 51.0/39\\{\footnotesize (without momentum resolution)} &&&&\\
\hline
\lw{data of Ref.~\cite{afan90}} & eq.~(\ref{eqn:a6}) & --- & $b = 1$
(fixed) & 53.9/37\\
& eq.~(\ref{eqn:a9}) & $2.96\pm 1.02$ & $b = 1$ (fixed) & 44.9/36\\
\hline\hline
data of Ref.~\cite{afan96} & eq.~(\ref{eqn:a6}) & --- & $0.43\pm
0.03$ & 55.0/46\\
8 mkm & eq.~(\ref{eqn:a9}) & $5.62\pm 3.03$ & $0.53\pm 0.07$ &
51.5/45\\
\hline
data of Ref.~\cite{afan96} & eq.~(\ref{eqn:a6}) & --- & $0.51\pm
0.04$ & 37.9/46\\
1.4 mkm & eq.~(\ref{eqn:a9}) & $4.44\pm 2.63$ & $0.60\pm 0.07$ &
35.1/45\\
\hline
\end{tabular}
\end{center}
\end{table}
%
%
\begin{table}[htbp]
\begin{center}
\caption[table2]{Results of the $\chi^2$ fits of $R(q)$ for
exponential source by eq.~(\ref{eqn:a9}).}
\medskip
\label{tbl:a2}
\begin{tabular}{c|cccc}
\hline
& $q_{\rms{lim}}$ [MeV] & $1/\beta$ [fm] & $b$ & $\chi^2/NDF$\\
\hline
8 mkm & 25 & $3.19\pm 3.55$ & $0.51\pm 0.08$ & 24.0/20\\
\hline
1.4 mkm & 25 & $3.00\pm 3.88$ & $0.59\pm 0.10$ & 20.3/20\\
        & 30 & $0.65\pm 2.62$ & $0.54\pm 0.08$ & 23.1/25\\
        & 35 & $2.08\pm 2.77$ & $0.57\pm 0.08$ & 27.2/30\\
        & 40 & $2.60\pm 2.85$ & $0.58\pm 0.08$ & 30.3/35\\
\hline
\end{tabular}
\end{center}
\end{table}
%
%
\newpage
\section*{{\large\bf Figure Captions}}
\begin{description}
  \item[Fig.~1. ] Flow chart of the present procedure for Gamow
factor and/or Coulomb wave function, which includes momentum
resolution by generating the Gaussian random numbers.
  \item[Fig.~2. ] Results of $\chi^2$ fit for 8 mkm target of
$p\;+ \mbox{Ta } \to \pi^+\pi^- + X$ reaction with $q > 3$ MeV/c by
eq.~(\ref{eqn:a6}).
  \item[Fig.~3. ] Results of the $\chi^2$ fit for Ref.~\cite{afan90}
of $p\;+ \mbox{Ta } \to \pi^+\pi^- + X$ reaction with $q > 3$ MeV/c
by eq.~(\ref{eqn:a9}).
  \item[Fig.~4. ] Results of the $\chi^2$ fits for $p\;+ \mbox{ Ta }
\to \pi^+\pi^- + X$ reactions with $q > 3$ MeV/c by
eq.~(\ref{eqn:a9}): $(a)$ 8 mkm target; $(b)$ 1.4 mkm target.
  \item[Fig.~5. ] Physical picture of the obtained source size based
on the assumed resonance effects.
\end{description}
\newpage
\begin{figure}[h]
\setlength{\unitlength}{1in}
\begin{picture}(5.0, 5.0)
\includegraphics{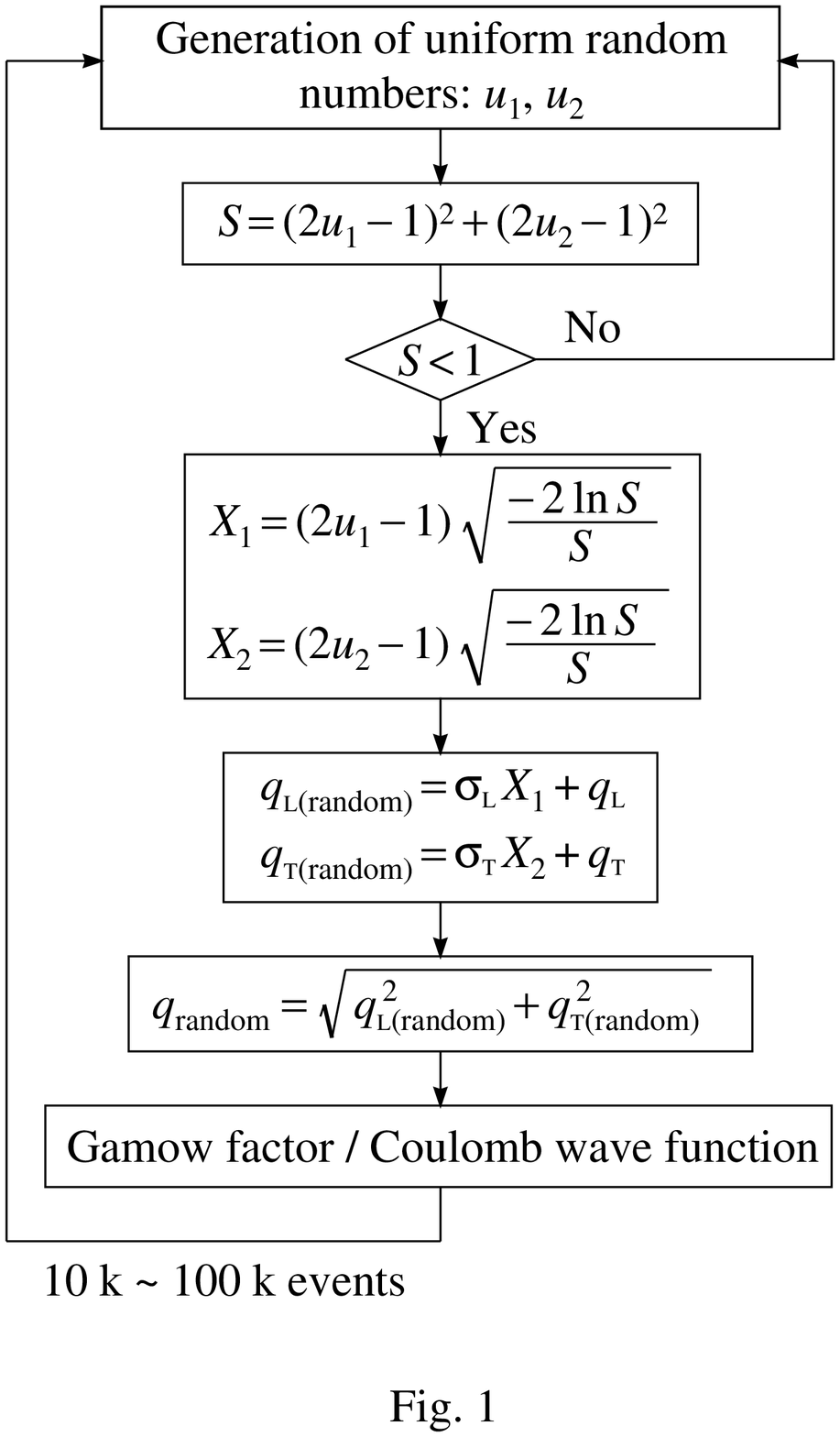}
\end{picture}
\end{figure}
\newpage
\begin{figure}[h]
\setlength{\unitlength}{1in}
\begin{picture}(5.0, 5.0)
\includegraphics{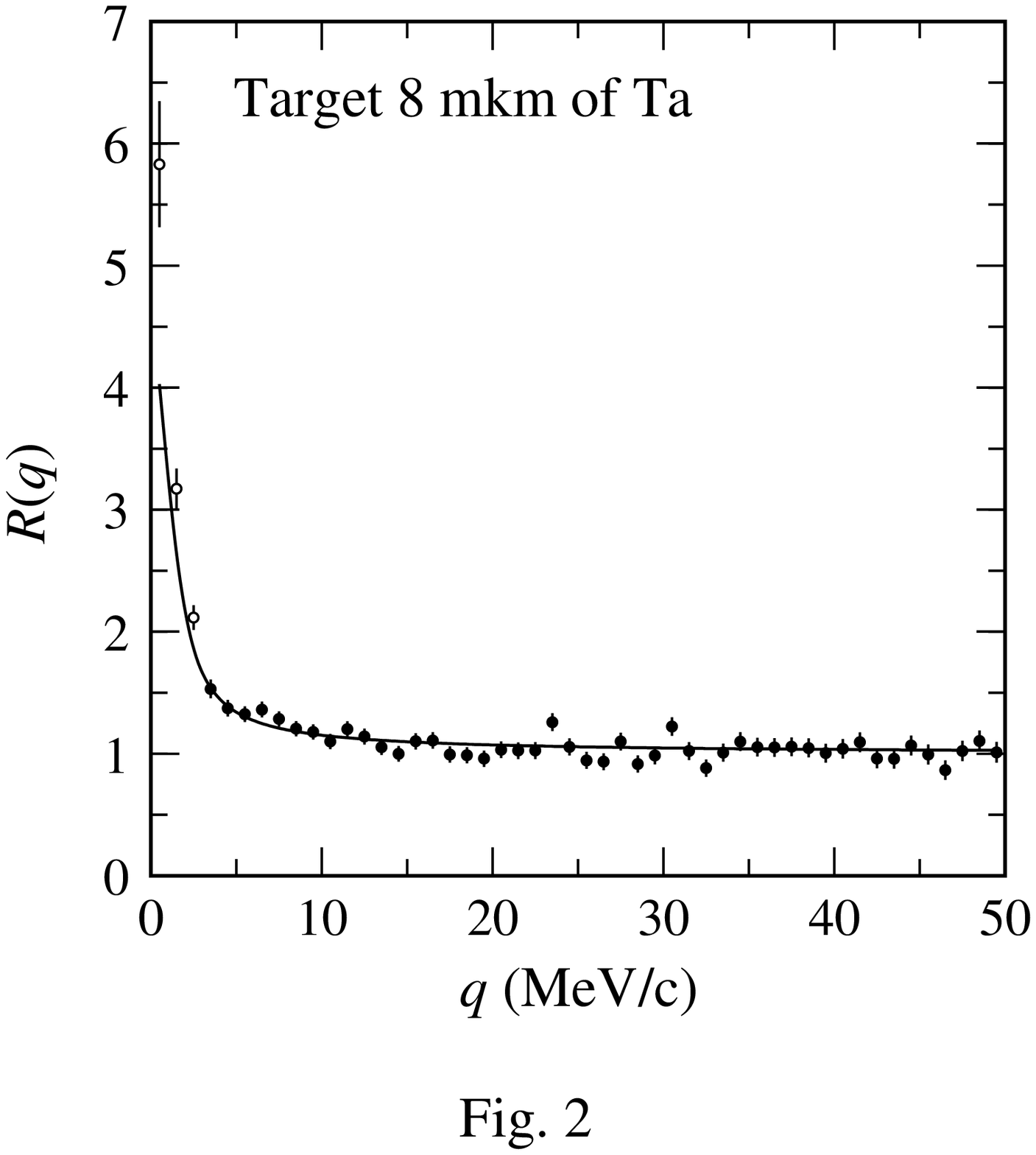}
\end{picture}
\end{figure}
\newpage
\begin{figure}[h]
\setlength{\unitlength}{1in}
\begin{picture}(5.0, 5.0)
\includegraphics{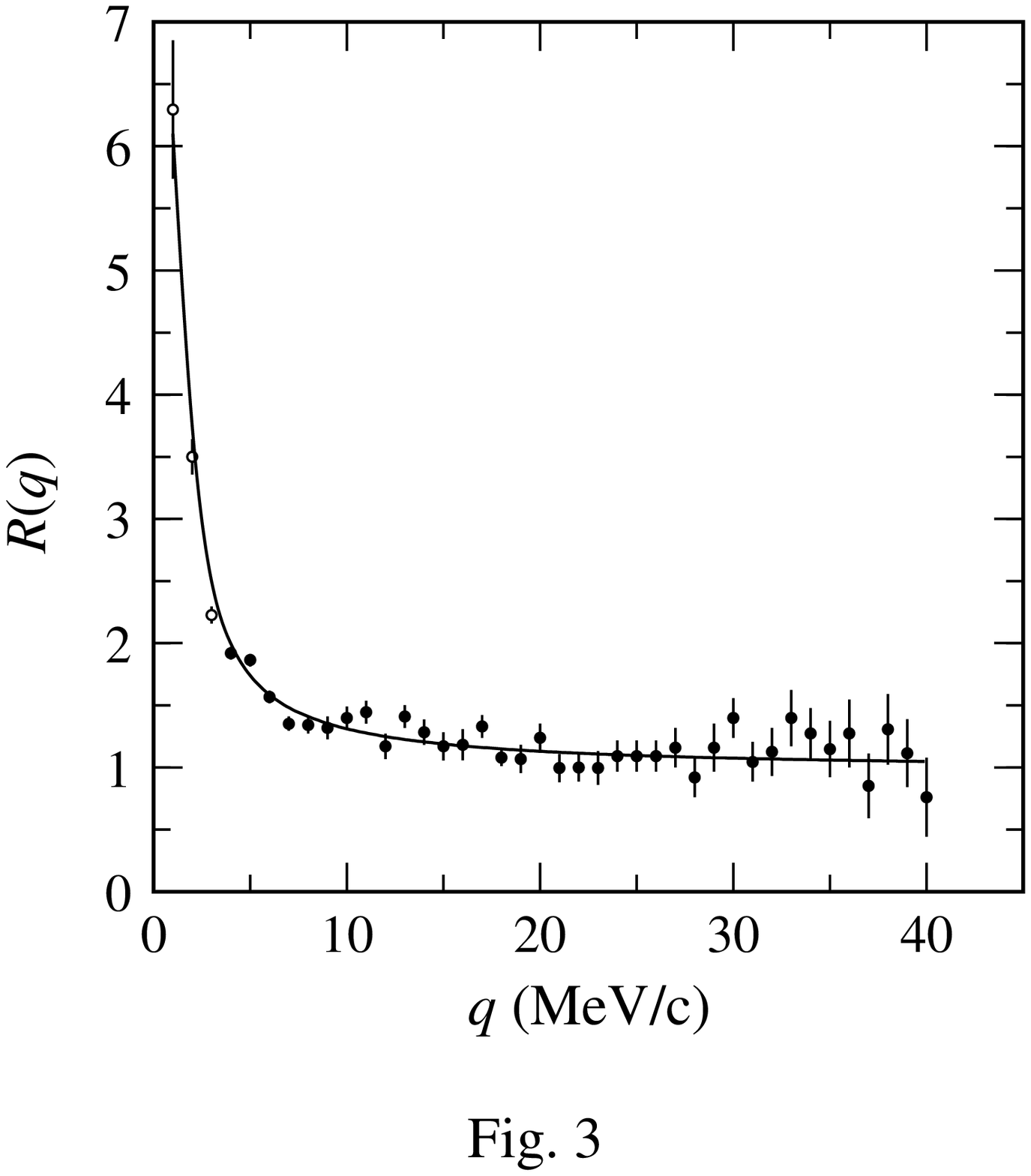}
\end{picture}
\end{figure}
\newpage
\begin{figure}[h]
\setlength{\unitlength}{1in}
\begin{picture}(5.0, 5.0)
\includegraphics{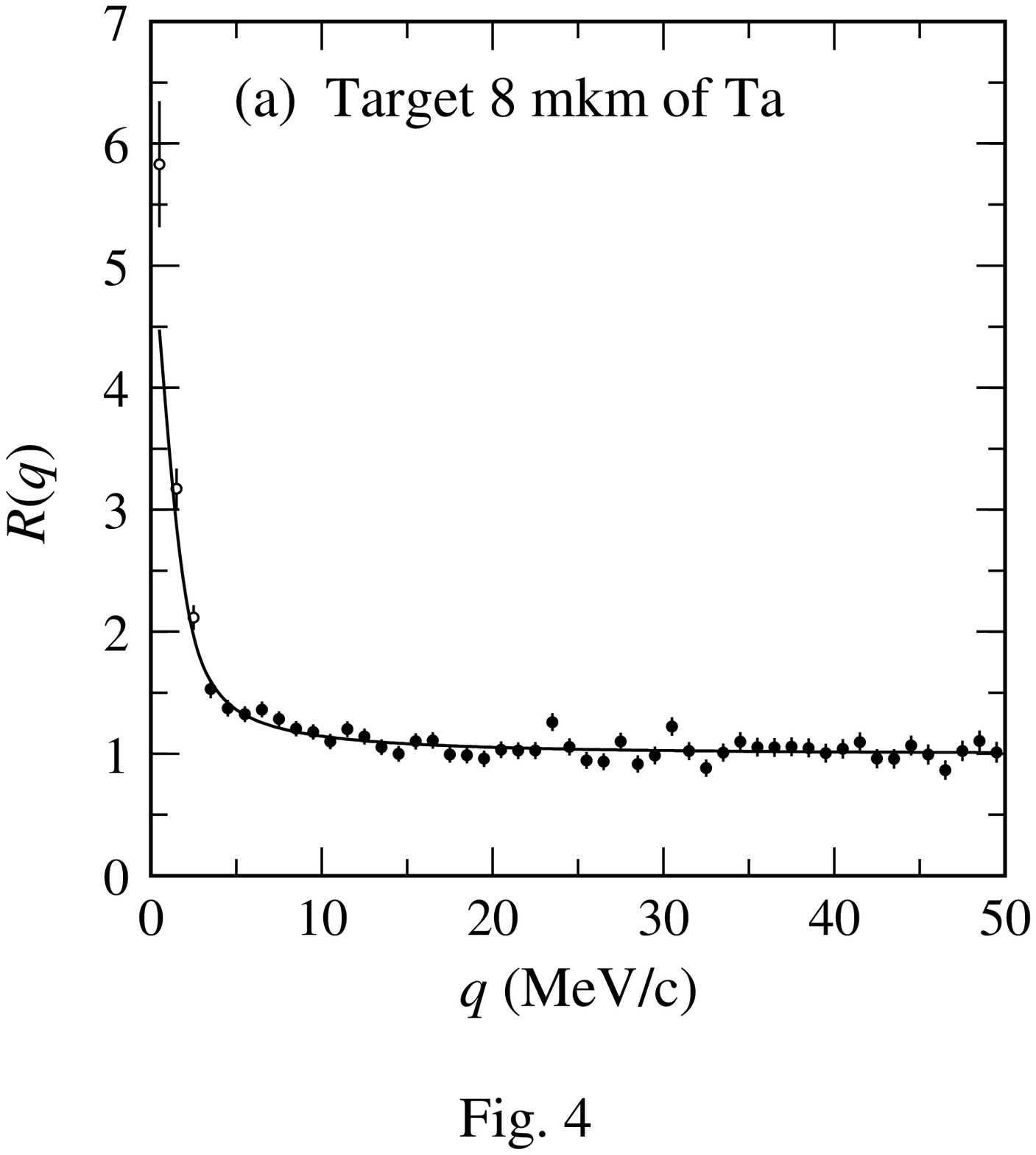}
\end{picture}
\end{figure}
\newpage
\begin{figure}[h]
\setlength{\unitlength}{1in}
\begin{picture}(5.0, 5.0)
\includegraphics{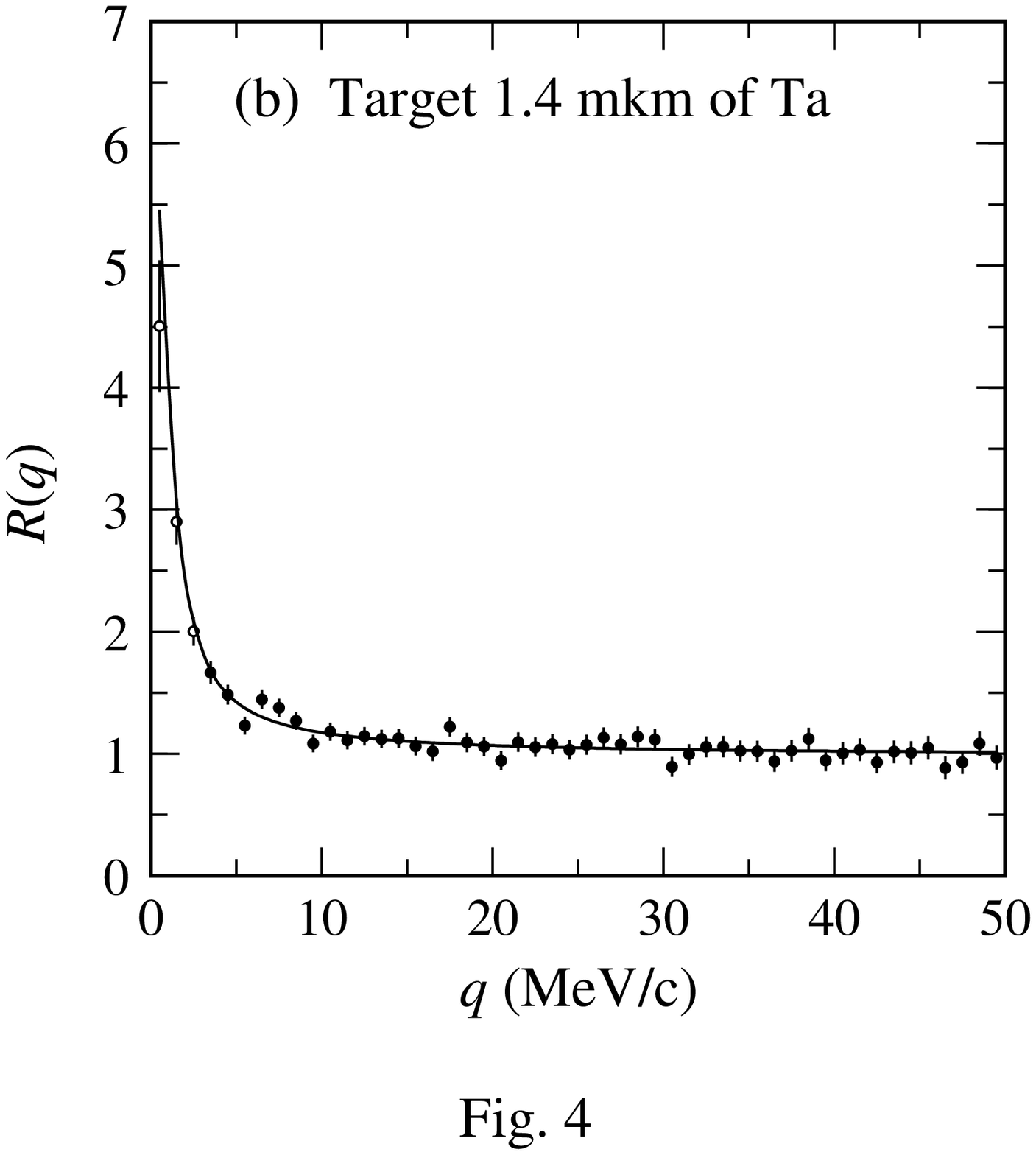}
\end{picture}
\end{figure}
\newpage
\begin{figure}[h]
\setlength{\unitlength}{1in}
\begin{picture}(5.0, 5.0)
\includegraphics{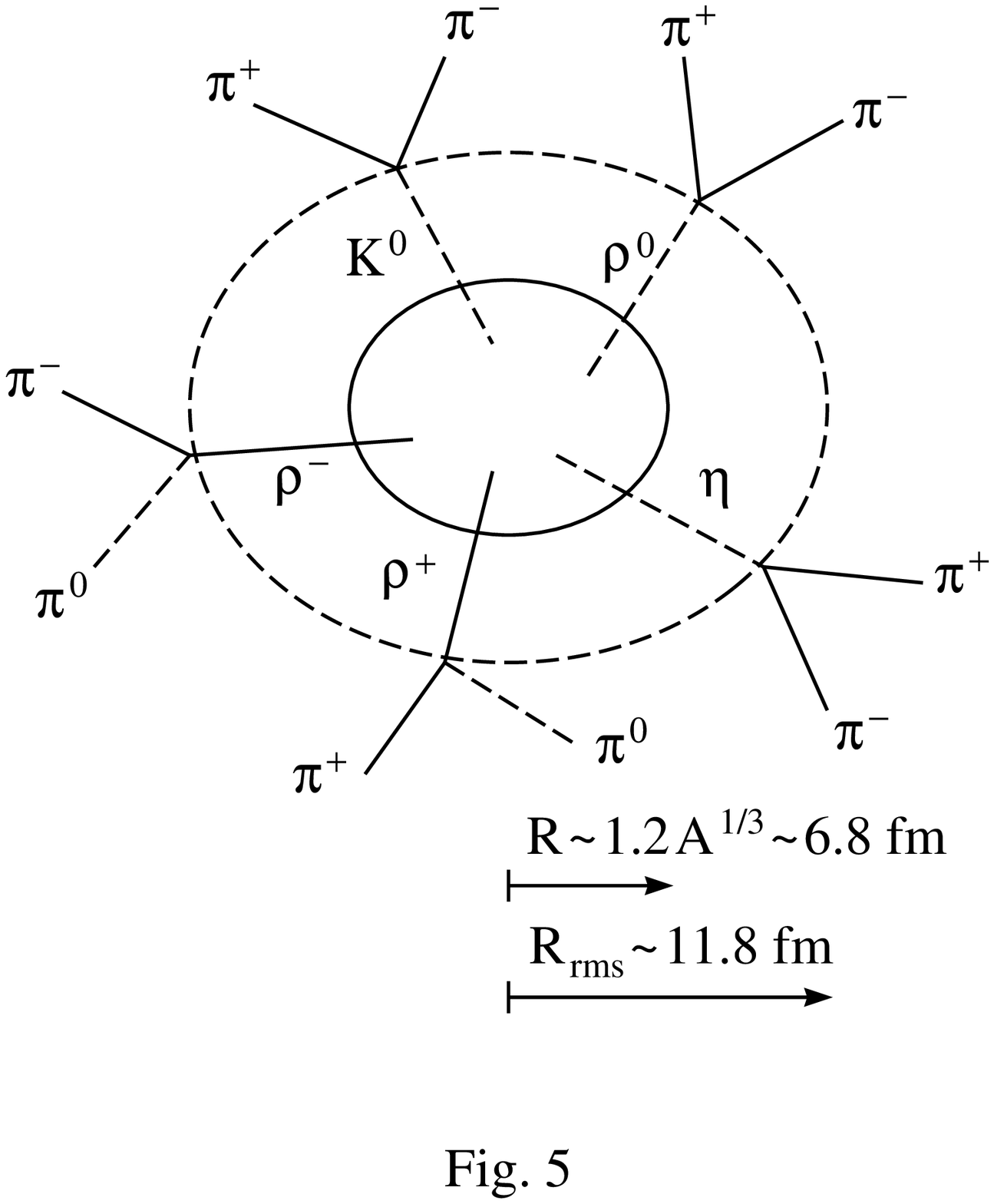}
\end{picture}
\end{figure}

\end{document}